\documentstyle[12pt]{article}

\newcommand {\PLB}[3]	{Phys. Lett.~{\bf B#1} (#2), #3}

\newcommand {\NPB}[3]	{Nucl. Phys.~{\bf B#1} (#2), #3}

\newcommand {\PRD}[3]	{Phys. Rev.~{\bf D#1} (#2), #3}

\newcommand {\PP}[3]	{Phys. Rep.~{\bf#1} (#2), #3}
\newcommand {\PRL}[3]	{Phys. Rev.~Lett.~{\bf#1} (#2), #3}
\newcommand {\JDG}[3]   {J. Diff.~Geom.~{\bf#1} (#2), #3}

\begin{document}
\begin{titlepage}
     \begin{normalsize}
     \flushright{UT-658}\\
     \end{normalsize}
\makebox[1in]{      }
\vfil
   \begin{LARGE}
       \begin{center}
       {Fermion Modes in Instanton-Anti-Instanton Background
       in (1+0)-dimensional Model}
       \end{center}
   \end{LARGE}
\vfil
% \vspace{2cm}
%
\begin{center}
    \begin{Large}
        Hajime Aoki
        \footnote{E-mail : haoki@tkyux.phys.s.u-tokyo.ac.jp}\\
    \end{Large}
         {\it  Department of Physics, University of Tokyo\\
               Bunkyo-ku, Tokyo, 113, Japan}\\
    \vfil
{\bf Abstract}
\par
\end{center}
We evaluate the eigenvalues of  Dirac operator in the background of
instanton-anti-instanton configuration (I-$\bar{{\rm I}}$), in a simple
(1+0)-dimensional model,
and find that quasi-zero-mode disappears for the closely localized
I-$\bar{{\rm I}}$ configurations.
This result suggests that the configurations which are dominant at high
energy in the
valley method are actually non-anomalous ones and irrelevant to fermion number
violation. Hence,
there seems to be no theoretical basis for
expecting anomalous cross section to become observable at energies of 10
Tev region in Weinberg-Salam theory.
\end{titlepage}
\vfil

\newpage

\section{Introduction}
\mbox{ }
\indent
It is  well known that in Weinberg-Salam theory baryon number is
violated due to anomaly non-perturbatively, but a naive semiclassical
approximation shows that such processes are strongly suppressed by the factor
of $e^{-2S_{{\rm I}}}$ where $S_{{\rm I}}$ is the instanton action
\cite{'t hooft}.
These processes had been therefore considered as
hopelessly unobservable until, in 1990, Ringwald \cite{ringwald} and Espinosa
\cite{espinosa} suggested that the cross section grows exponentially at high
energy when O(1/$\alpha$) gauge and Higgs bosons in the final states are
considered, because of the increase in available phase space.
Since then, a lot of modifications of their calculations have been made
\cite{review}, trying to solve the unitarity bound problem and to
apply them to the case at the energy above the Sphaleron.

Among them, the valley method \cite{balitsky yung,yung,khoze ringwald 1,khoze
 ringwald 2,aoyama} showed an optimistic result suggesting that the
anomalous cross
section reaches an observable one without breaking the unitarity bound
at the energy of the order of $E_{{\rm sp}}$ ($\sim$ 10Tev)
\cite{khoze ringwald 2}.
In this method, total cross section $\sigma _{tot}$ in the instanton
background
(I) is computed, via optical theorem, as the imaginary part of the forward
elastic scattering amplitude ($\Im (FES)$) in the instanton-anti-instanton
background(I-$\bar{{\rm I}}$),
\begin{equation}
\sigma _{tot,{\rm I}} = \Im (FES_{{\rm I}\mbox{-}\bar{{\rm I}}}),
\end{equation}
so that the final state corrections are automatically included.

However, since we are interested in the "anomalous" processes where baryon
number is violated, we have to extract the correct part from the
I-$\bar{{\rm I}}$
configurations, $i.e.$, fermion number of the intermediate state in the
background of I-$\bar{{\rm I}}$ configuration must be topologically correct,
and not
just the same as that of the initial (and final) state (Fig. 1).
When the initial state energy is increased, the configurations where the
separation between instanton and anti-instanton is much smaller than the
instanton size become dominant
\cite{khoze ringwald 1,khoze ringwald 2}, but such configurations seem to be
non-anomalous. Therefore the anomalous
part may be only a tiny fraction of their optimistic estimate of the
cross section \cite{konishi 1,konishi 2}.

Although both of the perturbative vacuum and the far separated
I-$\bar{{\rm I}}$
configuration have zero instanton number globally,
there should be some (qualitative) distinction between them.
In this paper, we consider a simple (1+0)-dimensional model, and investigate
the quasi-zero-mode of Dirac operator.

\section{(1+0)-dimensional Toy Model}
\setcounter{footnote}{0}
\mbox{}
\indent
In this section, we introduce a toy model which has the same features as the
Weinberg-Salam model, $i.e.$, instantons and anomaly.\footnote{Although
it doesn't appear to be possible to derive local anomaly
(non-vanishing current divergence) since it is a one dimensional model, it has
global anomaly as shown below.}
Closely related models have been used in a variety of contexts : Witten's
SUSY quantum mechanics \cite{witten} and fermion number fractionization
in polymer physics \cite{niemi}.
In all of these past applications, they considered quantum mechanics in
1-spatial dimension (i.e., (1+1)-dimensional field theory), whereas we use
this model as a (1+0)-dimensional field theory, and consequently physical
contents are completely different.

This model has a boson field $\Phi(t)$ in a double-well-potential and a
two-component fermion field $\Psi(t)$ coupled to the boson $\Phi(t)$ by
Yukawa-type interaction.
The Euclidean action and the associated Hamiltonian read as follows.
\begin{eqnarray}
S_{{\rm E}} & = & \int dt \, [\bar{\Psi} D \Psi
                     + \frac{1}{2} \dot{\Phi}^{2} + V(\Phi ) ], \\
D     & = & - i \sigma_{2} \partial_{t} + \sigma_{1} \Phi \nonumber \\
      & = & \left( \begin{array}{cc}
                0                &  - \partial_{t} + \Phi \\
             \partial_{t} + \Phi & 0
              \end{array} \right),\\
H     & = & \Phi \Psi^{\dagger} \sigma_{3} \Psi + \frac{1}{2} p^{2} + V(\Phi ),
\label{hamiltonian}
\end{eqnarray}
where $\sigma_{i}$'s are Pauli-matrices, and V($\Phi$) is the double well
potential.

First, let's consider the energy-spectrum of fermion when a background
configuration $\Phi$ is at the bottoms of the wells (Fig. 2(a)).
Since the Hamiltonian for fermion has the form in Eq. (\ref{hamiltonian}),
the up
(down)-component has
positive (negative) energy in the right well and negative (positive) energy
in the left, respectively (Fig. 2(b)). Therefore, when the background
configuration $\Phi$
is changed from the left well to the right well, the fermion number of up
(down) component increases by 1 ($-1$).

Next, let's consider the zero-mode of (1+0)-dimensional Dirac operator D in the
instanton (or anti-instanton) background.
The equations of zero-mode for up and down components
\begin{eqnarray}
   (\partial_{t} + \Phi ) \Psi_{u}  & = & 0 ,\\
 (- \partial_{t} + \Phi ) \Psi_{d}  & = & 0
\end{eqnarray}
can be easily integrated to give
\begin{eqnarray}
   \Psi_{u} & = & exp( -\int dt\, \Phi ), \\
   \Psi_{d} & = & exp( \int dt\, \Phi ).
\end{eqnarray}
When the background $\Phi$ is an instanton, $i.e.$
\begin{eqnarray}
  {\rm Index} & \equiv & \frac{1}{2} [{\rm sign} \Phi (t = \infty )
                         -{\rm sign} \Phi (t = - \infty) ]\label{index} \\
         &     =  & +1,
\end{eqnarray}
$\Psi_{u}$ is normalizable, but $\Psi_{d}$ is not.
In the case of anti-instanton the above situations are reversed.
Since D is a hermitian operator, $\Psi_{u}$ and $\bar{\Psi}_{d}$ have a
zero-mode in the instanton background,
while in the anti-instanton background $\Psi_{d}$ and $\bar{\Psi}_{u}$ have
a zero-mode. Therefore the following Green's functions give
non-vanishing results, in the instanton and anti-instanton background,
respectively,
\begin{eqnarray}
  \langle 0|T(\Psi_{u}(t_{1}) \bar{\Psi}_{d}(t_{2}))|0 \rangle & \neq & 0
   \quad (\mbox{instanton}),  \\
  \langle 0|T(\Psi_{d}(t_{1}) \bar{\Psi}_{u}(t_{2}))|0 \rangle & \neq & 0
   \quad (\mbox{anti-instanton}).
\end{eqnarray}
Consequently, the fermion numbers change as follows,
\begin{eqnarray}
  \triangle N_{u} & = & - \triangle N_{d} = 1 \quad (\mbox{instanton}), \\
  \triangle N_{u} & = & - \triangle N_{u} = -1 \quad (\mbox{anti-instanton}),
\end{eqnarray}
which gives the same results derived above from the energy-spectrum
consideration.

Finally, let's see the correspondence of non-zero-modes between up and down
components.
Dirac operator satisfies the following relations
\begin{eqnarray}
  \{ D , \sigma_{3} \}   & = & 0, \nonumber \\
  \left[ D^{2} ,\sigma_{3} \right] & = & 0,
\end{eqnarray}
where
\begin{eqnarray}
  D & = & -i \sigma_{2} \partial_{t} + \sigma_{1} \Phi
      =\left( \begin{array}{cc}
                0                &  - \partial_{t} + \Phi \\
             \partial_{t} + \Phi & 0
              \end{array} \right), \nonumber \\
  D^{2} & = & \left( \begin{array}{cc}
               - \partial_{t}^{2} + U_{u}        &  0 \\
                     0                           & - \partial_{t}^{2} + U_{d}
              \end{array} \right), \nonumber \\
  U_{u} & = & \Phi^2 - \partial_{t} \Phi ,\quad
  U_{d} = \Phi^2 + \partial_{t} \phi.
\end{eqnarray}
Consequently, the eigenstates for both $D^{2}$ and $\sigma_{3}$
\begin{eqnarray}
  D^{2} \Psi_{u} & = & \lambda^{2} \Psi_{u}, \quad
  \sigma_{3} \Psi_{u} = + \Psi_{u}, \\
  D^{2} \Psi_{d} & = & \lambda^{2} \Psi_{d}, \quad
  \sigma_{3} \Psi_{d} = - \Psi_{d},
\end{eqnarray}
have the following one to one correspondence for $\lambda \neq 0$,
\begin{eqnarray}
  \Psi_{d} & = & \frac{1}{\lambda} D \Psi_{u}, \\
  \Psi_{u} & = & \frac{1}{\lambda} D \Psi_{d}.
\end{eqnarray}
Therefore, we can obtain the eigenstates of D from these eigenstates of $D^{2}$
and
$\sigma_{3}$ in the following way,
\begin{equation}
  D \left[ \frac{\Psi_{u} \pm \Psi_{d}}{\sqrt{2}} \right]
  = \pm \lambda \left[ \frac{\Psi_{u} \pm \Psi_{d}}{\sqrt{2}} \right].
\end{equation}
\section{Quasi-Zero-Mode in I-$\bar{\mbox{I}}$ Background}
\mbox{}
\indent
Using the (1+0)-dimensional toy model presented in section 2, we
calculate numerically the spectrum of D in the background of
I-$\bar{{\rm I}}$ configurations,
\begin{equation}
  \Phi (t) = v[\tanh (\frac{1}{\rho} (t+\frac{1}{2} R))
               -\tanh (\frac{1}{\rho} (t-\frac{1}{2} R)) -1 ],
\end{equation}
where $v$ and $\rho$ are the vacuum expectation value of $\Phi$ and the
instanton size, respectively, which are determined by the double well
potential; R is the I-$\bar{{\rm I}}$ separation (Fig. 3).

Fig. 4 shows the smallest eigenvalue of $D^{2}$, and Fig. 5 to Fig. 7 show
their eigenfunctions.
These figures show that there exists a quasi-zero-mode whose eigenfunctions
are localized at the
position of instanton or anti-instanton for a large instanton-anti-instanton
separation R; this behavior suggests the existence of the
topologically correct fermions in the intermediate state.
On the contrary, when the distance divided by the instanton size, $R/\rho$,
becomes of the order of unity, the
quasi-zero-mode disappears and the transition to no-bound continuous mode
takes place; the intermediate state is no longer an anomalous one.

However, these figures also suggest that there is no definite value
of $R/\rho$ which
distinguishes the anomalous configurations from the non-anomalous ones.
Indeed the value of $R/\rho$ where the smallest eigen value departs from zero
appreciably, depends on the parameter $v$: the extension of the
eigenfunctions is of the order of $1/v$, and when they get close to overlap
each other, the smallest eigenvalue starts to increase.

\section{Conclusion and Discussion}
\mbox
\indent
We obtain a clear distinction between the perturbative vacuum and the
far separated instanton-anti-instanton configuration, when considered as
intermediate states in the valley method:
When instanton and anti-instanton get close and their separation becomes of the
order of instanton size,
the quasi-zero-mode of Dirac operator disappears.

Seen from Hamiltonian picture, we can obtain the same result.
The Hamiltonian for the up(down)-component fermion
is just the background $\Phi$ ($-\Phi$) in this model
(Eq. (\ref{hamiltonian})), and we can also see from Eq. (\ref{index})
that Chern-Simons number is also the value of background $\Phi$, essentially.
Consequently, it can be easily seen that the flow of
energy-spectrum crosses the zero value twice in the case of far separated
instanton-anti-instanton, whereas below the critical value
$R_{c} /\rho = 2\mbox{arctanh}(1/2) \sim 1.1$, the level crossings are absent.

Although we can obtain the clear results easily from the model which we have
presented in this paper,
this toy model is a much simplified
(1+0)-dimensional one, so some features could be different from those of a
realistic 4-dimensional SU(2) gauge theory.
For instance, in Ref. \cite{konishi 1} they claim that there's no
quasi-zero-mode
at all, nor is there any mode related to the fermion number violation
for I-$\bar{{\rm I}}$ background, in 4-dimensional gauge theory.

Nevertheless, since only the far separated instanton-anti-instanton
configurations which have correct topology in the intermediate states are
relevant for the baryon number violation, the optimistic
result based on the valley method seems to be invalid in the present toy
model, also.

The meaning of quasi-zero-mode in the Dirac operator and a precise analysis of
fermion number in
the intermediate states will be considered diagramatically elsewhere
\cite{aoki}.

\section*{Acknowledgement}
The author is grateful to Prof. K. Fujikawa, Prof. H. Kawai and Dr. S. Iso
for useful discussions and to Prof. K. Fujikawa and Dr. S. Iso for
a careful reading of the manuscript.

\newpage

\section*{Figure Captions}
Fig. 1.  Correct behavior of intermediate state fermions in the anomalous
         processes.\\
         At each vertex of instanton (I) and  anti-instanton
         ($\bar{{\rm I}}$), fermion number have to change correctly.\\[3ex]
Fig. 2.  Double-well-potential for boson (a), and energy-spectrum of
         fermions (b).\\[3ex]
Fig. 3.  Instanton-anti-instanton configuration.\\[3ex]
Fig.~4.  The smallest eigenvalue $\lambda^{2}$ of $D^{2}$ versus $R/\rho$,
         where
         R is the I-$\bar{{\rm I}}$ separation and $\rho$ is the instanton
         size.
         The vacuum expectation value of $\Phi$ at the bottoms of the wells,
         $v$, is set to be
         $1/(v\rho) = 2.5$ in this case.
         When $R/\rho$ becomes of the order of unity, quasi-zero-mode changes
         to continuous mode.
         Note that continuous modes start from $\lambda^{2} \sim v^{2}$.\\[3ex]
Fig. 5.  Wave function for $R/\rho = 15 $.\\
         Dashed line shows background $\Phi$, the straight line $\Psi_{u}$,
         and the dot line $\Psi_{d}$, respectively. When instanton and
         anti-instanton are far separated, wave functions of fermion are
         bounded at the instanton (or anti-instanton) position.\\[3ex]
Fig. 6.  Wave function for $R/\rho = 1.1$.\\
         Wave functions are overlapping.\\[3ex]
Fig. 7.  Wave function for $R/\rho = 0.1$.\\
         Dot-dashed line shows the wavefunction in the case of
         perturbative-vacuum.
         Transition from bound state to no-bound continuous mode
         takes place.

\end{document}